\documentstyle{elsart}
\input psfig
\def\lsim{\mathrel{\rlap{\lower4pt\hbox{\hskip1pt$\sim$}}
    \raise1pt\hbox{$<$}}}         
\def\gsim{\mathrel{\rlap{\lower4pt\hbox{\hskip1pt$\sim$}}
    \raise1pt\hbox{$>$}}}         
\begin{document}
\renewcommand{\topfraction}{1}
\renewcommand{\bottomfraction}{1}
\renewcommand{\textfraction}{.1}
\renewcommand{\floatpagefraction}{1}
\begin{frontmatter}

\title{What Will the First Year of SNO Show?}
\author{{John N. Bahcall}\thanksref{jnbemail}} 
\address{School of Natural Sciences, 
Institute for Advanced Study, Princeton, NJ 08540}
\author{Plamen I. Krastev\thanksref{pikemail}}
\address{Department of Physics, University of Wisconsin, Madison, WI 53706}
\author{Alexei Yu. Smirnov\thanksref{aysemail}}
\address{International Center for Theoretical Physics, 34100 Trieste, Italy}
\thanks[jnbemail]{E-mail address: jnb@ias.edu}
\thanks[pikemail]{E-mail address: krastev@nucth.physics.wisc.edu}
\thanks[aysemail]{E-mail address: smirnov@ictp.trieste.it}

\begin{abstract}
The ratio of the measured to the predicted standard model CC event
rates in SNO will be 0.47 if no oscillations occur.  The best-fit
active oscillation predictions for the CC ratio are: $0.35-39$ (MSW)
and $0.38-42$ (vacuum) (all for a $5$ MeV energy threshold), typically
about $20$\% less than the no-oscillation expectation.  We calculate
the predicted ratios for six active and sterile neutrino oscillation
solutions allowed at 99\% CL and determine the dependence of the
ratios on energy threshold.  If the high-energy anomaly observed by
SuperKamiokande is due to an enhanced $hep$ flux, MSW active solutions
predict that out of a total of $5000$ CC events above $5$ MeV in SNO
between $49$ and $54$ events will be observed above $13$ MeV whereas
only $19$ events are expected for no-oscillations and a nominal
standard $hep$ flux.
\end{abstract}

\end{frontmatter}

\section{Introduction}
The Sudbury Neutrino Observatory (SNO) has been designed to provide
definitive answers regarding neutrino properties that are manifested
in solar neutrino experiments~\cite{sno}.  

What will the first year of SNO show? 
We answer this question assuming 
that SNO will detect~\cite{sno} $\sim 3000$ to $5000$ neutrino
absorption (CC) events in the first year and that one
of the currently-favored neutrino oscillation solutions is
correct.

We focus on CC rate measurements that can be completed in the first
year of operation of SNO. Accurate measurements of the energy
spectrum, of time-dependences, and of the neutral current will require
more time.  We begin by determining in Sec.~\ref{sec:allowed} the
currently allowed MSW and vacuum oscillation solutions and by showing
that the no-oscillation hypothesis is rejected at a high CL even if
solar neutrino fluxes are treated as free parameters.  In
Sec.~\ref{sec:snovssuperk} we show that a ``smoking gun'' indication
of neutrino oscillations may be obtained by measuring the total CC
rate (neutrino absorption) in conjunction with the SuperKamiokande
measurement~\cite{superk} of the $\nu-e$ scattering rate.  This idea
has been discussed previously (see, e. g.,
refs.~\cite{sno,superk,response1}). What is new here is the
calculation of the expected range of the CC rate for each of the
currently favored oscillation solutions.  Figure~\ref{fig:compare}
summarizes our most important results.  In Sec.~\ref{sec:highenergy}
we calculate the number of high energy ($> 13$ MeV) CC events that are
expected if the high-energy anomaly observed by
SuperKamiokande~\cite{superk} is due to an enhanced $hep$ flux.  The
discussions in refs.~\cite{sno,superk,response1,bl,bari,response3}
describe other things that can be learned by comparing in detail the
results of the SuperKamiokande and SNO measurements.

\begin{table}[!b] 
\centering 
\caption[]{\label{tab:bestfit} {\bf Best-fit global oscillation
parameters and confidence limits.} 
The active neutrino solutions are from Fig.~\ref{fig:global}. 
The differences of the squared masses are given in ${\rm eV^2}$.  
}
\begin{tabular}{lccc} 
\noalign{\bigskip}
\hline 
\noalign{\smallskip}
Scenario&$\Delta m^2$&$\sin^2(2\theta)$&C.L. \\
\noalign{\smallskip}
\hline\hline 
\noalign{\smallskip}
 LMA&  $2.7\times10^{-5}$ &$7.9\times10^{-1}$&$68$\% \\
 SMA& $5.0\times10^{-6}$&$7.2\times10^{-3}$& $64$\%\\
 LOW& $1.0\times10^{-7}$ &$9.1\times10^{-1}$&$83$\%\\ 
 ${\rm VAC_S}$ & $6.5\times10^{-11}$ &$7.2\times10^{-1}$ & $90$\%\\
 ${\rm VAC_L}$ & $4.4\times10^{-10}$ &$9.0\times10^{-1}$ &$95$\% \\
 ${\rm Sterile}$ & $4.0\times10^{-6}$ &$6.6\times10^{-3}$&$73$\% \\
\noalign{\smallskip}
\hline
\noalign{\smallskip}
\end{tabular}
\end{table}

\section{Allowed solutions}
\label{sec:allowed}

\begin{figure}[!ht]
\centerline{\psfig{figure=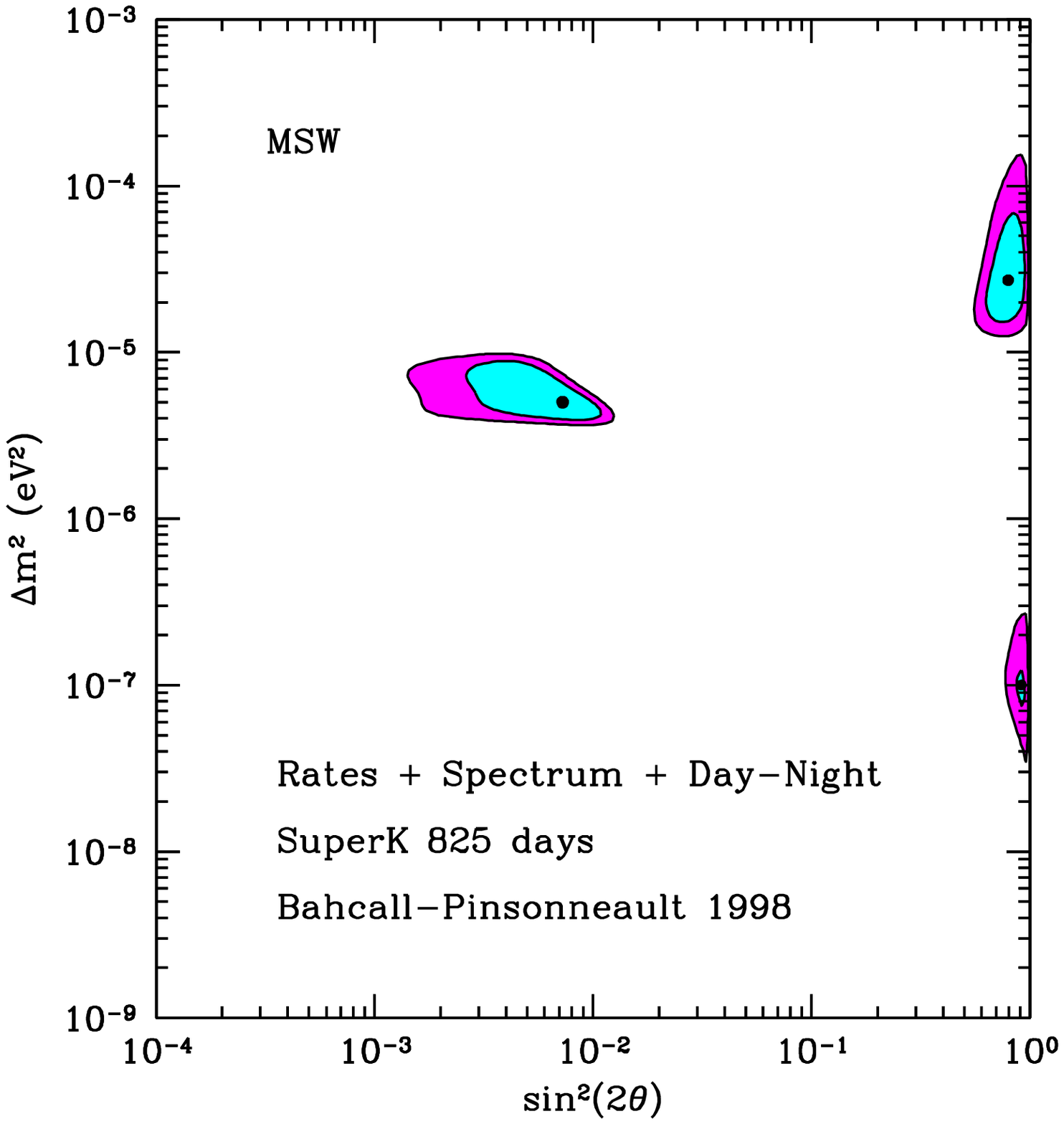,width=3.5in}}
\vglue-.6in
\centerline{\psfig{figure=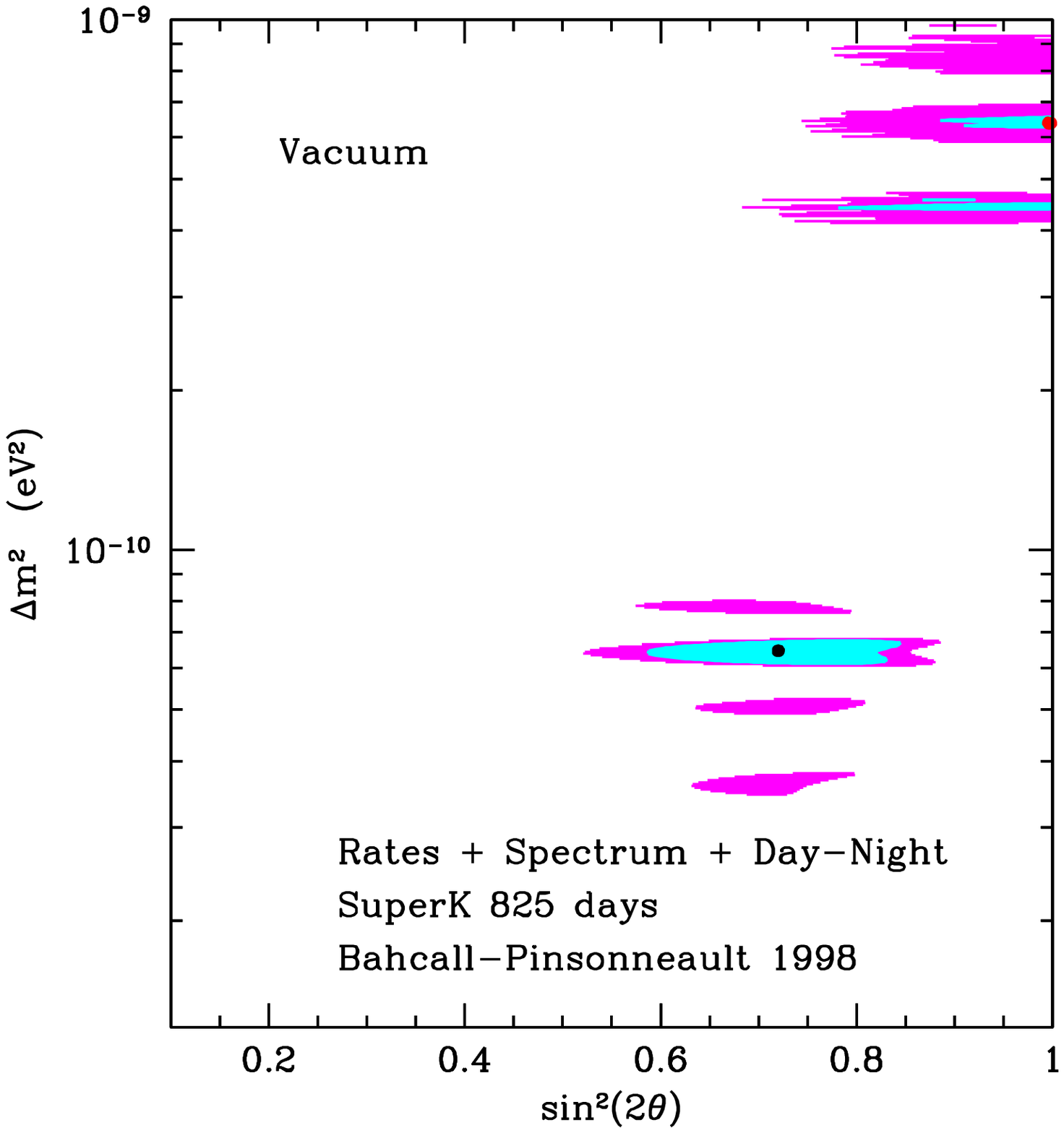,width=3.5in}}
\vglue-.4in
\caption[]{\small Global oscillation solutions.
The input data include the total rates in the Homestake,
Sage,  Gallex, and 
SuperKamiokande experiments, as well as the electron recoil energy
spectrum and the Day-Night effect measured by SuperKamiokande in 825
days of data taking.
Figure~\ref{fig:global}a shows the
global solutions for the allowed MSW oscillation regions,
known, respectively, as
the SMA, LMA, and LOW solutions~\cite{bks98}.
Figure~\ref{fig:global}b shows
the global solution for the allowed vacuum oscillation regions.
The CL  contours correspond, for both panels, to 
$\chi^2 = \chi^2_{\rm min} + 4.61 (9.21)$, representing  
90\% ( 99\% CL) relative to each of the best-fit solutions (marked by
dark circles)  given in
Table~\ref{tab:bestfit}. The best vacuum fit to the
SuperKamiokande electron recoil energy spectrum is
marked in Fig.~\ref{fig:global}b at $\Delta m^2 = 6.3 \times10^{-10}$ eV$^2$ 
and $\sin^2 2\theta = 1$.
\label{fig:global}}
\end{figure}

Figure~\ref{fig:global} shows the allowed global solutions for
MSW~\cite{msw}  and
vacuum~\cite{vac} two-neutrino oscillations. 
Table~\ref{tab:bestfit} gives the best-fit parameters and confidence
limits (C.L.) for the
different neutrino scenarios, including sterile neutrinos (not shown
in Fig.~\ref{fig:global} since the allowed region is similar to the
SMA region,\cite{bks98}).
The experimental  data include
the total rates for the Homestake (chlorine)
experiment~\cite{chlorine}, 
the SAGE~\cite{sage} and
GALLEX~\cite{gallex} (gallium) experiments, and the SuperKamiokande (water
Cherenkov)~\cite{superk} experiment. 
We also include the SuperKamiokande~\cite{superk}
electron recoil energy
spectrum and the difference between the
average day and night event rates. 
We follow the procedures described in our earlier work on global solar
neutrino solutions~\cite{bks98}.
The vacuum solutions that correspond to a mass smaller than
$10^{-10}{\rm eV^2}$,${\rm  VAC_{S}}$  , 
fit somewhat better to the total rates (ignoring
the spectrum measurement) than the larger-mass solutions, ${\rm  VAC_{L}}$. 
The spectrum is better fit by the larger-mass solutions.
The ${\rm  VAC_{L}}$ solutions, but not the ${\rm  VAC_{S}}$
solutions,  predict large seasonal variations for
the $^7$Be and pp neutrinos. The predicted seasonal and day-night
effects for the $^8$B neutrinos are small for all of the oscillation 
scenarios.

Is there any way, even very artificial, of avoiding the conclusion
that some new  physics is
required to explain the solar neutrino experimental results?
We adopt the maximally skeptical attitude and ignore
everything we have learned about the sun and about solar models over
the past four decades. 
We allow~\cite{bks98,robertson} 
the $pp$,
${\rm ^7Be}$, ${\rm ^8B}$, and CNO  fluxes to take on any non-negative
values, requiring only that the shape of the continuum spectra be
unchanged (as demanded by standard electroweak theory).
The sole constraint on the fluxes is the requirement that the
luminosity of the sun be supplied by nuclear fusion reactions among
light elements (see section 4 of Ref.~\cite{howwell}); we 
use the standard
value for the essentially 
model-independent  ratio of $pep$ to $pp$ neutrino fluxes. 

The search for the best fit fluxes yields:
${\rm ^7Be/(^7Be)}_{\rm SM} = 0.0$,
${\rm ^8B/(^8B)}_{\rm SM} = 0.47$, $pp/(pp)_{\rm SM} = 1.10$, and
CNO/${\rm CNO_{\rm SM} = 0.0}$, where SM stands for the combined
standard solar~\cite{bp98} and electroweak model. 
These values were found by an extensive
 computer search
with four free parameters and the luminosity constraint.
The minimum value of $\chi^2$ is, for 3 d.o.f.,

\begin{equation}
\chi^2_{\rm min} \hbox{( {\rm five rates + D-N effect};
~arbitrary $pp$, ${\rm ^7Be}$, ${\rm ^8B}$; CNO)} = 13.8\ .
\label{chinoCNO}
\end{equation}

There are no acceptable solutions at the $99.7$\% C. L. ($3\sigma$). 
We  carried out searches using a variety of other
prescriptions: {\it ab initio} setting the CNO fluxes equal to zero,
combining the two gallium experiments, 
and including or excluding the Homestake~\cite{chlorine} or 
Kamiokande~\cite{kamiokande} experiments.
The  poor
fit is robust. In all cases, there are no acceptable solutions at a CL
of about $3 \sigma$ or higher. If new physics is required to describe
solar neutrino propagation, then this result can be strengthened in
the future by adding new results from SNO.

\section{SNO versus SuperKamiokande}
\label{sec:snovssuperk}

Let  
\begin{equation}
R ~\equiv~ \frac{{\rm Observed~Rate}}{{\rm Standard~Model~Rate}}
\label{eq:defnr}
\end{equation}
be the dimensionless ratio of the observed rate (in any experiment)
divided by the rate expected from the combined standard solar and 
electroweak model(SM).
The experimental result for SuperKamiokande after 825 days of data
taking 
with a 6.5 MeV energy threshold is~\cite{superk}
\begin{equation}
R_{\rm SK} = 0.475 \pm 0.015 .
\label{eq:rskexp}
\end{equation}

The neutrino-electron scattering experiments, Kamiokande and
SuperKamiokande, contain contributions from both charged current and
neutral current interactions. Hence, for SuperKamiokande the ratio
of observed to standard rates can be written
\begin{eqnarray}
\!\!\!\!\!\!\!\!\!\!\!
R_{\rm SK}\!\! =\!\! \frac{f_B \int dE_e \epsilon(E_e) \int dE \phi_{\rm SM} (E)
\left[\sigma_{\nu_e-e}(E,E_e) 
P(E)\! +\! \sigma_{\nu_\mu-e}(E,E_e) (1\!\! -\!\!
P(E))\right]}{\int dE_e \epsilon(E_e) 
\int dE\phi_{\rm SM}(E)\sigma_{ \nu_e-e}(E,E_e)} ,\nonumber\!\!\!\!\!\\
\label{eq:rsktheo}
\end{eqnarray}
where 
$\phi_{\rm SM}(E)$ is the standard model $^8$B neutrino flux, $P(E)
= P(E, \nu_e \rightarrow \nu_e)$, $\sigma_{\nu_e-e}$ and
$\sigma_{\nu_\mu-e}$ are the scattering cross sections~\cite{sirlin},
$E_e$ is the electron energy, and $\epsilon(E_e)$ is the integrated
resolution and efficiency function, and 
$f_B$ is defined by 

\begin{equation}
f_B ~=~ { {\phi(^8B)_{\rm true}} \over {\phi(^8B)_{\rm SM}} }.
\label{eq:defnt}
\end{equation}

Neutrino absorption in SNO is a pure charged-current reaction.
Therefore, 
\begin{equation}
R_{\rm CC, SNO} = \frac{f_B \int dE_e \epsilon(E_e) 
\int dE \phi_{\rm SM} (E)\left[\sigma_{{\rm
abs}} (E,E_e) P(E) \right]}{\int dE_e \epsilon(E_e) 
\int dE\phi_{\rm SM}(E)\sigma_{{\rm abs}}(E,E_e)},
\label{eq:rsnotheo}
\end{equation}
where $\sigma_{\rm abs}$ is the absorption cross section.
Equation~(\ref{eq:rsktheo}) and Eq.~(\ref{eq:rsnotheo}) are 
independent of the total flux $\phi_{\rm SM}$.
In intermediate calculations, we use the Bahcall-Pinsonneault 1998
value for the standard model $^8$B flux: $5.15\times10^{6}{\rm
cm^{-2} s^{-1} }$~\cite{bp98}. 
For the no-oscillation hypothesis, $f_B$ is
determined by SuperKamiokande measurements, $f_{B, \rm no
~ oscillation} = 0.475$.

The standard electroweak model, which embodies the no neutrino
oscillation hypothesis, predicts that 

\begin{equation}
R_{\rm SK} ~=~ R_{\rm CC, SNO},
\label{eq:nooscil}
\end{equation}
independent of the SuperKamiokande and SNO energy thresholds.
Equation~(\ref{eq:nooscil}) is valid because, in the absence of
new electroweak physics like oscillations, the only theoretical 
reason for  $R_{\rm SK}$ or $ R_{\rm CC, SNO}$
being  different from unity is that the standard solar model neutrino
flux is wrong, i. e., $f_B \not= 1$. Since $R_{\rm SK}$ and $ R_{\rm CC, SNO}$
are both proportional to $f_B$, Eq.~(\ref{eq:nooscil}) is correct
independent of the calculated standard model flux.

Table~\ref{tab:ratecc} gives the calculated values of $R_{\rm CC,SNO}$
that are predicted by all of the allowed neutrino oscillation regions
shown in Fig.~\ref{fig:global} as well as the sterile neutrino
solution (see Sec.~\ref{sec:allowed}).  At each allowed point, the
best-fit ratio, $f_B$, of observed to SM neutrino flux was determined
by minimizing the $\chi^2$ fit of the calculated $\nu-e$ electron
recoil energy spectrum with respect to the measured SuperKamiokande
recoil energy spectrum\footnote{The uncertainties in the theoretical
predictions for $R_{\rm CC,SNO}$ have been reduced by a factor of
between three and four compared to our pre-SuperKamiokande 1996
calculations for LMA, SMA, and VAC (see Table VII of
ref.~\cite{howwell}).} .  The survival probabilities $P(E)$ are
determined by the neutrino oscillation parameters. The neutrino
absorption cross sections were evaluated using the computer routine
(with speed-up modifications) provided by Bahcall and Lisi.  We used
all three cross section evaluations in the Bahcall-Lisi
subroutines~\cite{bl}; the theoretical uncertainty in $R_{\rm CC,SNO}$
due to the cross section calculations is $\pm 3$\% , much smaller than
the spread due to the uncertainty in neutrino parameters.  We use the
preliminary SNO collaboration estimates for the energy resolution,
absolute energy scale, and detection efficiencies~\cite{bl}. Detailed
calculations show that uncertainties in these quantities affect the
rates by $\sim $ $1$\% or $2$\%, which is small compared to the total
range of the oscillation predictions. Systematic uncertainties ($\sim
3$\% for SuperKamiokande) and precise values for the SNO detector
characteristics must be determined by measurements with SNO and by
detailed Monte Carlo simulations.

\begin{table}[!t] 
\centering 
\caption[]{\label{tab:ratecc} {\bf Predicted
charged current rate in the SNO detector.}  The table shows the
predicted ratio, $R_{\rm CC,SNO}$, 
 of the measured $\nu_e$ absorption rate to the combined standard
model rate for the different neutrino scenarios shown in
Fig.~\ref{fig:global} and for sterile neutrinos (similar to the global
SMA solutions).
For no oscillations,
$R_{\rm CC,SNO} = R_{\rm SK} = 0.47$.
The uncertainties indicated result from the variation
of the neutrino parameters within the
90\% CL (and the 99\%
CL) globally-allowed regions of Fig.~\ref{fig:global}.
The second, third, and fourth columns
were evaluated using a threshold of 5 MeV, 7 MeV, or 8 MeV, respectively, for
the total electron energy.
The last column gives the range of $f_B$
found at $90$\% and $99$\% CL by fitting to the measured 
SuperKamiokande recoil
energy spectrum at each globally-allowed set
of oscillation parameters.}
\begin{tabular}{lcccc} 
\noalign{\bigskip}
\hline 
\noalign{\smallskip}
Scenario&$R_{\rm CC,SNO}$ &
$R_{\rm CC,SNO}$&$R_{\rm CC,SNO}$&$f_B$\\
Threshold&5 MeV&7 MeV&8 MeV&\\
\noalign{\smallskip}
\hline\hline 
\noalign{\smallskip}
 No osc.& $0.475 \pm 0.015 $ &$0.475 \pm 0.015 $&$0.475 \pm
0.015$&$0.475 \pm 0.015$ \\[7pt]
 LMA& $0.35_{-0.05(-0.06)}^{+0.01(+0.05)}$ &$0.35_{-0.05(-0.06)}^{+0.01(+0.04)}$ &$0.35_{-0.05(-0.06)}^{+0.01(+0.04)}$&$1.140^{0.905(0.875)}_{1.325(1.450)}$\\[7pt]
 SMA& $0.39_{-0.04(-0.07)}^{+0.03(+0.07)}$ &$
0.42_{-0.03(-0.06)}^{+0.01(+0.05)}$&$0.43_{-0.03(-0.06)}^{+0.01(+0.05)}$&
$ 0.955^{0.62(0.555)}_{1.23(1.290)}$\\[7pt]
 LOW& $0.38_{-0.03(-0.03)}^{+0.02(+0.02)} $ & $0.38_{-0.03(-0.03)}^{+0.02(+0.02)}$  & $0.38_{-0.03(-0.03)}^{+0.02(+0.02)}$&$0.995^{0.910(0.885)}_{0.995(1.15)}$\\[7pt] 
${\rm VAC_S}$&$0.38^{+0.04(+0.08)}_{-0.06(-0.07)}$&$ 0.42^{+0.02(+0.06)}_{-0.05(-0.09)}$&$0.44^{+0.02(+0.05)}_{-0.04(-0.10)}$&$ 1.005^{0.815(0.600)}_{1.315(1.425)}$\\[7pt]
${\rm VAC_L}$&$ 0.42^{+0.02(+0.03)}_{-0.01(-0.04)}$&$0.41^{+0.03(+0.04)}_{-0.02(-0.03)}$&$ 0.40^{+0.04(+0.07)}_{-0.03(-0.05)}$&$ 0.785^{0.710(0.690)}_{0.845(0.830)}$\\[7pt] 
${\rm MSW, Sterile}$& $0.48^{+ 0.005(+0.01)}_{-0.005(-0.01)}$&  $0.50^{+0.02(+0.03)}_{-0.02(-0.03)}$&$0.50^{+0.04(+0.05)}_{-0.01(-0.02)}$&$0.930^{0.630(0.550)}_{1.245(1.415)}$\\[7pt]
\noalign{\smallskip}
\hline
\noalign{\smallskip}
\end{tabular}
\end{table}

\begin{figure}[!ht]
\centerline{\psfig{figure=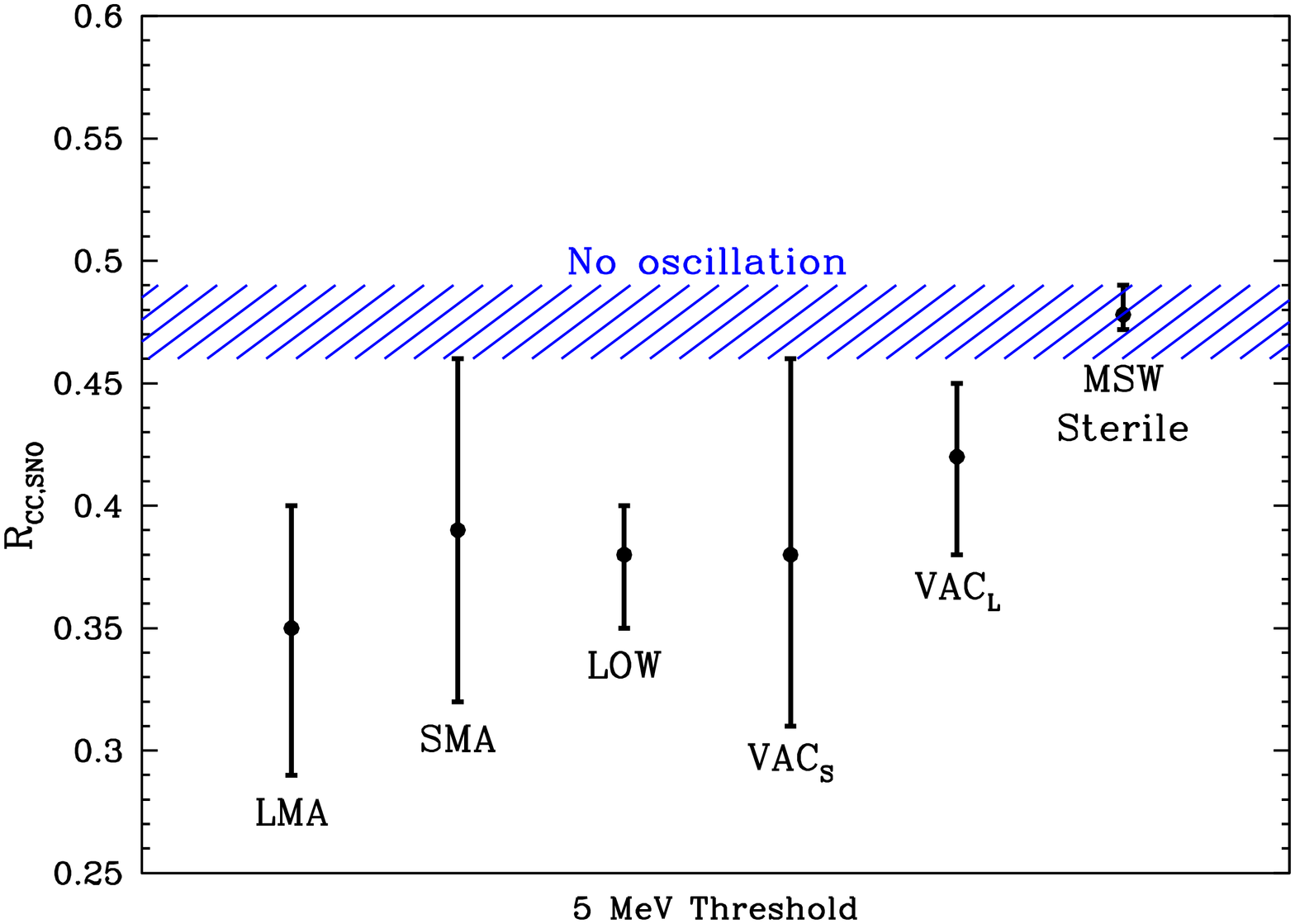,width=4in,angle=270}}
\centerline{\psfig{figure=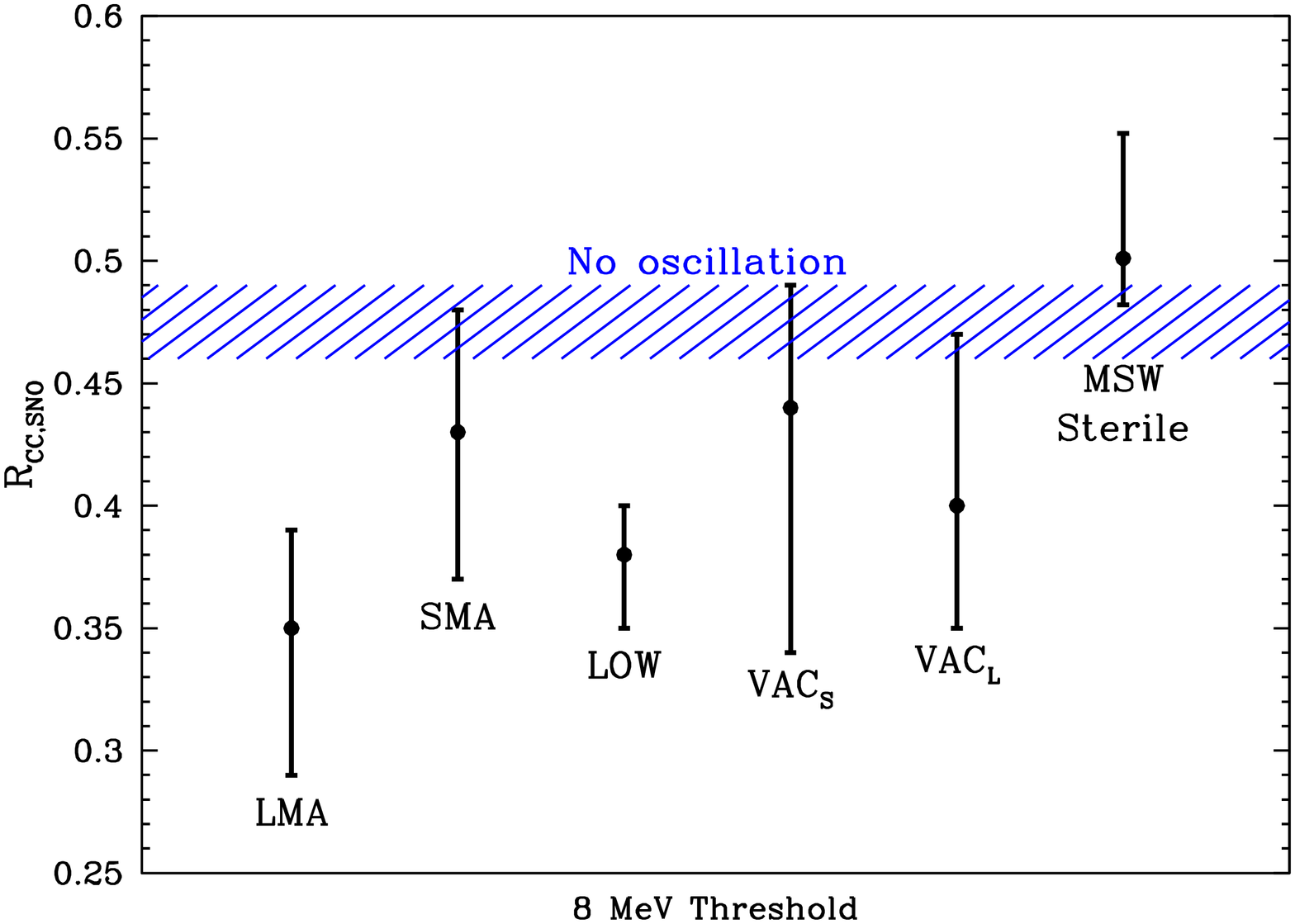,width=4in,angle=270}}
\caption[]{\small Comparison of the CC SNO rate  and the
no oscillation prediction.
The shaded area is  the no oscillation prediction based upon the 
measured SuperKamiokande rate for
$\nu-e$ scattering, see
Eq.~(\ref{eq:rskexp}) and Eq.~(\ref{eq:nooscil}). 
The SNO CC ratios, Eq.~(\ref{eq:rsnotheo}), 
are shown on the vertical axes for
different neutrino scenarios and two different total electron energy
thresholds, 5 MeV and 8 MeV. The error bars on the neutrino
oscillation results
represent the range of values predicted by the 99\% CL allowed neutrino
oscillation solutions displayed in Fig.~\ref{fig:global}. 
\label{fig:compare}}
\end{figure}

Figure~\ref{fig:compare} shows visually the predicted results when
$R_{\rm CC, SNO}$ is compared with $R_{\rm SK}$.  The error bars in
Fig.~\ref{fig:compare} correspond to the $99$\% CL global solution
regions of Fig.~\ref{fig:global}.  The statistical uncertainties after
one year of CC measurements with SNO are expected to be $\sim 2$\%,
which is much less than the range in the neutrino predictions.

For oscillations into active neutrinos, Fig.~\ref{fig:compare} and
Table~\ref{tab:ratecc} show that $R_{\rm CC, SNO}$ should be
significantly different from $R_{\rm SK}$ unless the correct solution
lies in a relatively small region of the allowed parameter space for
neutrino oscillations.  The average value of the best-fit solutions
listed in Table~\ref{tab:ratecc} is $R_{\rm CC, SNO} = 0.38$, which is
about $20$\% less than no-oscillation value of $0.475$.

We have evaluated the ratios $R_{\rm CC, SNO}$ for different
thresholds in order to exhibit the dependence of the ratios on the
threshold energy.  A 5 MeV threshold is one of the goals of the SNO
collaboration and an 8 MeV threshold would essentially eliminate
background from the neutral current reaction, as well as decreasing
other troublesome backgrounds~\cite{sno}.  For the LMA and LOW
solutions, Fig.~\ref{fig:compare} and Table~\ref{tab:ratecc} show that
there is no significant advantage in using a lower threshold, but for
the SMA and $VAC_S$ solutions there is a substantial improvement in
discriminatory power by lowering the threshold from 8 MeV to 5 MeV.

The results for sterile neutrinos are important.  One might initially
guess that the sterile neutrino solution would give a result very
similar to the no-oscillation result. This is the case for a threshold
of 5 MeV (see top panel of Fig.~\ref{fig:compare}). However, for an 8
MeV threshold the sterile neutrino solution predicts $R_{\rm CC, SNO}$
that are generally larger than $R_{\rm SK}$ (bottom panel of
Fig.~\ref{fig:compare}). The reason is that survival probability for
the SMA solution is an increasing function of neutrino energy in the
region of interest~\cite{bks98}.  If the SMA or sterile neutrino
solution is correct, then the 8 MeV threshold measurement will sample
on average a higher survival probability than the SuperKamiokande
measurement performed with a $6.5$ MeV threshold.  In the event that
initial measurements performed with an $8$ MeV threshold yield a value
of $R_{\rm CC,SNO}$ that is close to or slightly larger than $0.48$,
then a measurement of the shape of the recoil electron energy spectrum
to as low an energy as possible will be an important test of the
predicted small energy dependence of the survival probability of the
MSW Sterile solution.

\section{The high-energy anomaly in SNO}
\label{sec:highenergy}

The recoil energy spectrum measured by SuperKamiokande shows
evidence for an enhanced event rate above a total electron energy
of $13$ MeV~\cite{superk} . Several possible explanations for this
anomaly have been suggested, including:
1) an enhanced flux of the high energy $hep$
neutrinos~\cite{bk98,frere};
2) a real upturn in the survival probability that
is described by vacuum neutrino oscillations~\cite{superk}; 3) a statistical
fluctuation~\cite{superk}; and 
4) a systematic error  in the absolute energy
calibration~\cite{bks98}. The first two explanations imply that a high-energy
anomaly will also be observed in SNO, while the third and fourth
explanations suggest that SNO will not show a high-energy anomaly.

Because of the good intrinsic energy resolution of the deuterium
reaction, SNO can discriminate well among these possibilities. To
illustrate the sensitivity to the high-energy anomaly, we have
computed electron recoil energy spectra in the SNO detector using the
best-fit MSW neutrino oscillation scenarios illustrated in
Fig.~\ref{fig:global} with the $hep$ flux treated as a free parameter.

If the $hep$ flux is equal to the nominal SM value and the electron 
recoil energy spectrum is not distorted, 
then  $19$ out of a total of $5000$ CC events are expected
to be above $13$ MeV\footnote{If there are no oscillations  but the
$hep$ flux (as well as the  $^8$B flux) is 
allowed to vary within $99$\% CL, then
there are between $36-64$ ($13-29$) CC events above $13$ ($14$) MeV,
corresponding to an $hep$ range between $6$ and $23$ times the nominal
SM flux.}.
Since the SuperKamiokande measurements have shown a recoil electron
spectrum that is not significantly distorted below $13$ MeV, 
possibilities 3) and 4) also predict 
about $19$ CC events above $13$ MeV in SNO.
The expected number of high energy events is very different if the $hep$ flux
is enhanced or if vacuum oscillations cause the high-energy anomaly
(possibilities 1 and 2 above).

For different globally-acceptable neutrino oscillation scenarios, 
Table~\ref{tab:hep} shows the total number of CC events expected above
$13$ MeV and $14$ MeV out of a total of 5000 CC events.
For a threshold of $13$ MeV,
the difference between the $hep$-enhanced and the non-enhanced energy
spectrum is more than $7\sigma$ for the LMA and LOW solutions and about
$4.4\sigma$ for the SMA solutions.
The two vacuum oscillation regions predict a large range of
high-energy events, some of which are not much above the nominal
SM solution.

\begin{table}[!ht] 
\centering 
\caption[]{\label{tab:hep} {\bf High-energy events.}  
The table shows the number of events predicted
at large energies out of a total of $5000$ CC events.
For all the oscillation scenarios,  the $hep$ flux
was calculated by evaluating the best-fit value for $\phi({\rm
hep})/\phi({\rm hep})_{\rm SM}$ at representative points in the $99$\%
CL globally allowed regions shown in Fig.~\ref{fig:global}.
For the no-oscillation case, the standard $^8$B spectrum shape 
was fit to the SuperKamiokande result with the 
$hep$ flux fixed at the nominal SM value.
}
\begin{tabular}{lccc} 
\noalign{\bigskip}
\hline 
\noalign{\smallskip}
Scenario&Events above&Events above&$\phi({\rm hep})/\phi({\rm hep})_{\rm SM}$ \\
&~~~~$13$ MeV &~~~~$14$ MeV & \\
\noalign{\smallskip}
\hline\hline 
\noalign{\smallskip}
 No osc.& 19&4&1\\
 LMA&  $51-54$&$22-24$& $27-47$ \\
 SMA& $49-53$&$18-23$& $17-19$\\
 LOW& $50-51$ &$21-22$&$26-35$\\ 
 ${\rm VAC_S}$ & $32-54$ &$6-23$ & $0-40$\\
 ${\rm VAC_L}$ & $22-73$ &$3-25$ &$0-31$ \\
 ${\rm MSW \, Sterile}$ & $30-43$ &$$8 - 18 &$13-25$ \\
\noalign{\smallskip}
\hline
\noalign{\smallskip}
\end{tabular}
\end{table}

\section{Discussion: Discovering smoking guns}
\label{sec:discussion}

The standard electroweak model predicts that essentially nothing
happens to neutrinos after they are created in the center of the
Sun. Solar neutrinos should all be $\nu_e$ (produced by beta-decay of
proton rich elements) and to high accuracy ($1$ part in $10^5$) the
$^8$B solar neutrino energy spectrum should have the same shape as the
laboratory energy spectrum~\cite{b8spectrum}. Hence, if the standard
electroweak model is correct, the ratio, $R_{\rm CC,SNO}$, of the
measured neutrino absorption rate in SNO to the rate predicted by the
combined standard solar and electroweak model must equal the ratio,
$R_{\rm SK}$, of the measured rate in SuperKamiokande to the rate
predicted by the combined standard model.

Any departure from this equality, Eq.~(\ref{eq:nooscil}), would be a
``smoking gun'' indication of new physics. Figure~\ref{fig:compare}
and Table~\ref{tab:ratecc} show that SNO has a reasonable chance of
discovering this smoking gun in the first year of operation.  The mean
of the best-fit solutions for oscillating into active neutrinos yields
a mean ratio $R_{\rm CC,SNO}$ of $0.38$, which is $20$\% less than the
no-oscillation expectation of $0.475$.  If a smoking-gun is to be
discovered, Nature must be reasonably cooperative and not choose an
extreme oscillation solution, i. e., an SMA or vacuum solution that
produces one of the largest values of $R_{\rm CC,SNO}$ that are allowed by
the existing global solutions.  If either SMA or vacuum oscillations
is the correct solution, then the chances of discovering a smoking gun
violation of Eq.~(\ref{eq:nooscil}) would be improved significantly by
using a lower threshold like 5 MeV rather than a more easily
obtainable threshold like 8 MeV (see Fig.~\ref{fig:compare} and
Table~\ref{tab:ratecc}).  The predictions of the LMA and the LOW
solutions are well separated from the measured value of $R_{\rm SK}$
for thresholds between 5 MeV and 8 MeV (cf. Fig.~\ref{fig:compare}).

It will be very difficult to discriminate between different
oscillation solutions involving active neutrinos using just
measurements of the CC rate. Figure~\ref{fig:compare} shows that the
different oscillation solutions predict largely overlapping values of
$R_{\rm CC,SNO}$.

Sterile neutrinos predict values for $R_{\rm CC,SNO}$ that are
generally well separated from the solutions for active neutrinos and
which show a significant dependence upon energy threshold.
For an energy threshold like 8 MeV, the
sterile solutions predict values for $R_{\rm CC,SNO}$ that are
even larger than the no-oscillation solution (see
Fig.~\ref{fig:compare}). 

For the six neutrino oscillation scenarios summarized in
Table~\ref{tab:ratecc}, 
the best-fit total  $^8$B flux ranges between $0.79$ and $1.14$ of the
standard solar model flux~\cite{bp98}. The globally-allowed solutions
span, at $99$\% CL, 
 a total range between $0.55$ and $1.32$ of the standard model
flux, comparable to the $3\sigma$ model uncertainty.

The high-energy anomaly observed by SuperKamiokande above 13 MeV
could be a smoking gun indication of vacuum oscillations~\cite{superk}
or it may be due to a relatively large flux of $hep$
neutrinos~\cite{superk,bk98} or to observational
factors~\cite{superk,bks98}. 
Table~\ref{tab:hep} summarizes the number of high-energy neutrino
events predicted by different neutrino oscillation scenarios. The 
currently favored MSW solutions
predict that between $38$ and $74$ events, out of a total of $5000$ CC
events, should be observed above
$13$ MeV if there is an enhanced $hep$ flux. Only $19$ out of $5000$
CC events should be above  $13$ MeV if the standard electroweak model
is correct and the $hep$ is equal to its nominal SM value.
Vacuum oscillations allow a wide range, 
between $24$ and $62$, of higher-energy events.

The results of the first year of operation of SNO will be exciting.

We are grateful to colleagues who made suggestions that improved
the initial draft of this paper.
JNB  acknowledges support from NSF grant No. PHY95-13835 
and
PIK acknowledges support from  NSF grant No. PHY95-13835  and
NSF grant No. PHY-9605140. Much of the work for this paper was done
while JNB was a visitor in the physics department 
at the Weizmann Institute of Science and PIK was attending the
`Neutrino Summer' program at CERN.

\end{document}